\documentclass[twocolumn,aps,amssymb, prl]{revtex4}
\usepackage{epsfig}
\usepackage{graphicx}

\begin{document}

\title{Reentrant temperature dependence of critical current in superconductor
- ferromagnet - superconductor junctions based on PdFe alloys.}
\author{V.V. Bol'ginov, O. M. Vyaselev, V.N. Shilov}
\email{ryazanov@issp.ac.ru}

\affiliation{Institute of Solid State Physics, Russian Academy of
Sciences, Chernogolovka, 142432, Russia}

\begin{abstract}
The magnetic and transport properties of $Pd_{0.99}Fe_{0.01}$ thin
films have been studied. We have found that the Curie temperature
of the films is about 20 K and the magnetic properties strongly
depend on temperature below $T_{Curie}$. We have also fabricated
the set of superconductor-ferromagnet-superconductor josephson
junctions $Nb-PdFe-Nb$. The temperature dependence of the
junctions with the ferromagnet layer thickness of about 36 nm
shows the reentrant behaviour that is the evidence of the
transition of the junction into the $\pi$-state.
\end{abstract}

\maketitle


The proximity effect in superconductor-ferromagnet heterostructures
attracts a great interest of several recent decades. As it was shown
earlier (~\cite{Buz82,Buz92} and \cite{Buz2005} as a review) the
decay of the superconducting order parameter in the ferromagnet is
accompanied by the sign-changing oscillations. The unambiguous
evidence of sign-reversal spatial oscillations of the
superconducting order parameter in a ferromagnet was provided by the
observation of the $\pi$-state in $SFS$ Josephson junctions
\cite{PRL,Aprili,Sellier,JLTP}. '$\pi$-junctions' \cite{Bul77} are
weakly coupled superconducting structures with the ground state
phase difference of the macroscopic superconducting wavefunction
$\varphi =\pi$. They are characterized by the anomalous
current-phase relation $I_s=I_c\sin (\varphi + \pi) =-I_c\sin
\varphi $ with negative critical current~\cite{Bul77}. Spatial
oscillations of the superconducting order parameter in a ferromagnet
close to an $SF$-interface were predicted in Ref.~\cite{Buz92}. The
physical origin of the oscillations is the exchange splitting of
spin-up and spin-down electron subbands in ferromagnets. The period
is $\lambda_{ex}=2\pi \xi_{F2}$, where the oscillation (or
"imaginary") length $\xi_{F2}$ can be extracted from the complex
coherence length $\xi_F$ in a ferromagnet: $\frac 1 {\xi_{F}}=\frac
1{\xi_{F1}}+i\frac 1 {\xi_{F2}}$. In the case of large exchange
energy and negligible magnetic scattering in the $F$-layer the
imaginary length $\xi_{F2}$ and the order parameter decay length
$\xi_{F1}$ are equal \cite{Buz92}: $\xi_{F1}=\xi_{F2}=\sqrt{\hbar
D/E_{ex}}$, where $D$ is the diffusion coefficient for electrons in
a ferromagnet and $E_{ex}$ is the exchange energy responsible for
sign-reversal superconductivity in a ferromagnet. The transition
into the $\pi$-state manifests itself as a reentrant behaviour of
critical current density as a function of F-layer thickness.

The most promising way to observe the transition to the $\pi$-state
is to use weak ferromagnetic alloys as an interlayer. Weak exchange
interactions in such alloys lead to large values of characteristic
lengths: the oscillations period $\lambda_{ex}$ and the decay length
$\xi_{F1}$. This gives an opportunity to obtain a detailed
$j_c(d_F)$ curve and to demonstrate its reentrant behaviour (see
\cite{SFS-06} for CuNi alloys and \cite{Aprili} for PdNi alloys).
Additionally weak ferromagnetism in alloys allows to drive 0-pi
transition point by temperature (\cite{PRL}, \cite{SFS-06}). This
causes the reentrant temperature dependence of SFS junctions
critical current density that is the unambiguous proof of the
$0-\pi$ transition. This effect was observed for a first time in
Ref. \cite{PRL} in $Nb-Cu_{0.48}Ni_{0.52}-Nb$ SFS-junctions and
later the reentrant $I_c(T)$ dependence was observed in
SFS-junctions based on others $Cu_{1-x}Ni_{x}$ alloys
(\cite{Sellier}, \cite{SFS-06}, \cite{Weides}).

\begin{figure}[tbp]
\begin{center}
\includegraphics[width=0.45\textwidth, clip]{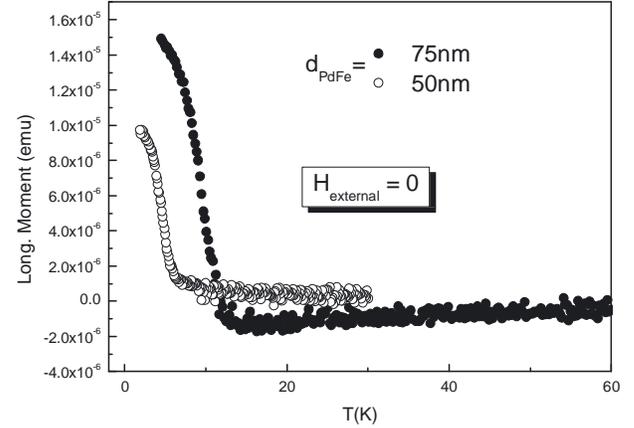}
\end{center}
\caption{The magnetic moment vs temperature dependance of thin film
$Pd_{0.99}Fe_{0.01}$ in zero external magnetic field obtained by
means of SQUID-magnetometry.}\label{SQUID-magn}
\end{figure}

It is well known that alloys $Cu_{1-x}Ni_x$ possess the
ferromagnetic properties only at sufficiently large concentrations
$x>0.44$. The large amount of ferromagnetic impurities is able to
modify strongly the transport properties of initial materials and to
complicate an analysis of an experimental data. One can avoid this
difficulties if using the alloys with very small concentration of
ferromagnetic atoms ($x\leq0.01$). For realization of this approach
it is very useful to use Pd or Pt as a starting nonmagnetic
material. According to stoner criteria this materials are very close
to ferromagnetic ordering and insertion of very small amount of
magnetic atoms is to lead to strong polarization of electrons. If
the amount is large enough and the polarized regions overlap then
the alloy will possess ferromagnetic properties. The ferromagnetic
properties of alloys $Pd_{1-x}Fe_{x}$ were studied in Refs.
\cite{114}-\cite{127}.

\begin{figure}[tbp]
\begin{center}
\includegraphics[width=0.45\textwidth, clip]{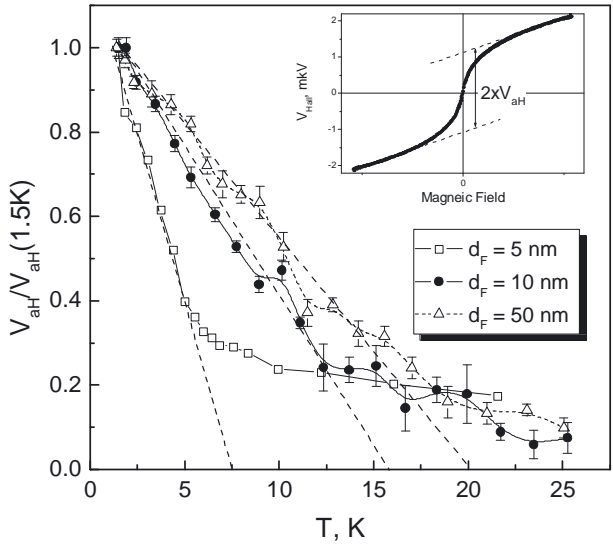}
\end{center}
\caption{The normalized Hall voltage vs temperature dependance of
thin film $Pd_{0.99}Fe_{0.01}$. Dashed lines show an extrapolation
of this curves to zero voltages in order to estimate the Curie
temperature. The inset shows the typical Hall voltage vs the
magnetic field dependance and the definition of $V_{aH}$ value.}
\label{Hall}
\end{figure}

In this article we have studied the magnetic and transport
properties of thin films $Pd_{1-x}Fe_x$ obtained by rf-sputtering in
argon atmosphere. The concentration of Fe in target was about 1\%.
The geometry of samples was formed by means of rf-ion etching in
argon plasma. The resistivity of films in thickness range from 5 to
50 nm was about 40 $\mu\Omega\cdot cm$ and very slightly differed
with thickness and temperature. The magnetic properties of PdFe
films were studied by means of SQUID-magnetometry and the anomalous
Hall effect measurements. We found that spontaneous magnetic moment
arises at temperatures below $10\div12~K$ if the film thickness is
greater than 50 nm (see fig. \ref{SQUID-magn}). We did not find any
magnetic response studying the thinner films. Hall measurements
showed that the ferromagnetic properties of the films increases
smoothly as the temperature decreases below 20-25 K (see fig.
\ref{Hall}). We observed no sufficient hysteresis on $V_{aH}$ vs
magnetic field curves up to temperatures as small as 1.5 K. From the
experimental data we can estimate that the Curie temperature of the
films with $d_{PdFe} < 50 nm$ is about $T_{Curie}~=~15\div 20~K$.
Note that for the film of very small thickness (5 nm) $T_{Curie}$ is
much smaller - about 7.5 K. This effect may be due to either worse
structure of first atomic layers of the film or due to size effect
as the film thickness of 5 nm is of the order of distance between
$Fe$ atoms.

\begin{figure}[tbp]
\begin{center}
\includegraphics[width=0.45\textwidth, clip]{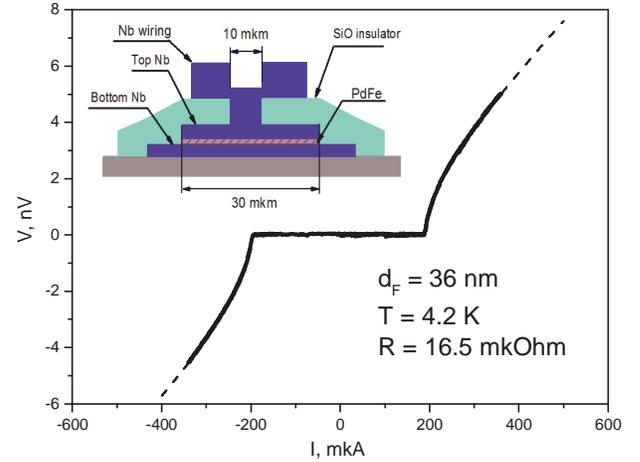}
\end{center}
\caption{Typical $IV$ curve for josephson SFS-junctions
$Nb-Pd_{0.99}Fe_{0.01}-Nb$. The dashed curve on the main panel shows
the fit of the experimental data in order to determine the junction
resistance. Inset shows a schematic cross section of the junction.}
\label{IV}
\end{figure}

Josephson SFS sandwiches Nb-PdFe-Nb were preparing in 5 steps. First
of all we sputtered three layers (bottom Nb layer, PdFe layer and
top Nb layer) in the same vacuum cycle that ensured the high
transparency of SF-interfaces. On the second step we formed junction
area by means of photolithography, chemical etching of top Nb layer
and ion etching of intermediate PdFe layer. The junction size was
about 30x30 mkm$^2$. The bottom Nb layer was not damaged during the
chemical etching of top Nb as it was protected by chemical stable
$PdFe$ layer. On the third step we formed the geometry of the bottom
Nb electrode using photolithography and chemical etching. The fourth
step consisted of the sputtering of isolating SiO layer by means of
thermal evaporation and lift-off photolithography. The SiO layer
thickness was about 300 nm and the window size was 10x10 mkm$^2$. On
the last step we made the Nb wiring of about 400 nm thickness by
means of magnetron sputtering. The geometry of top electrode was
formed by lift-off photolithography. The sample was subjected to ion
etching before the sputtering to ensure good superconducting contact
between the top Nb electrode of the junction and the Nb wiring. The
cross-section of the junction is shown on inset in fig. \ref{IV}.

\begin{figure}[tbp]
\begin{center}
\includegraphics[width=0.45\textwidth, clip]{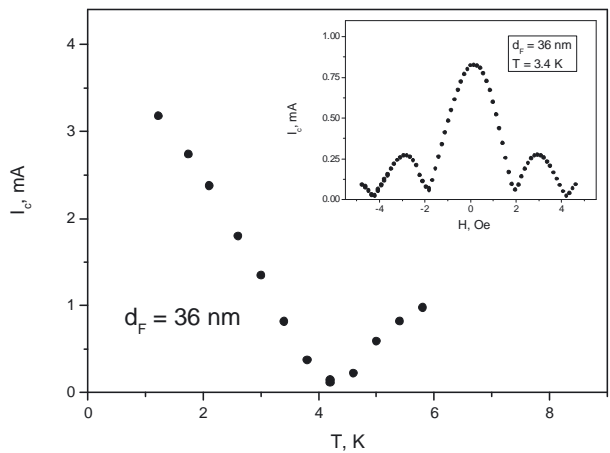}
\end{center}
\caption{The reentrant critical current vs temperature dependance of
$Nb-Pd_{0.99}Fe_{0.01}-Nb$ junction with the ferromagnetic layer
thickness 36 nm. Inset shows the critical current vs temperature
dependance of the junction.}\label{IT}
\end{figure}

The IV-curves of our samples can be well described by the well known
equation $V~=~R\sqrt(I^2-I_c^2)$ that allows to estimate the sample
resistance as a fitting parameter. The approximation of experimental
curve shown in fig. \ref{IV} gives $R \approx 16.5 \mu\Omega$. We
assume that the junction resistance is the sum of interlayer
resistance $\rho d_F/A$ and the boundary resistance $R_B/A$, where
$\rho = 40 \mu\Omega\cdot cm$ is the resistivity of PdFe,
$d_F~=~36~nm$ interlayer thickness, $A = 9\cdot 10^{-6}~cm^2$ -
junction area, $R_B$ - the boundary resistance per unit area. So we
can estimate that the $R_B = 4.5\cdot10^{-12} \Omega\cdot cm^2$ and
the transparency parameter \cite{KuLu} is equal
$\Gamma_B=R_B/AR=0.03$. The small value of $\Gamma_B$ indicates the
high transparency of the SF-interfaces. The $I_c(H)$ dependance
shown on inset in fig. \ref{IT} has the form of Fraughofer pattern
$$I_c(\Phi) \sim \frac{\sin{\pi\Phi/\Phi_0}}{\pi\Phi/\Phi_0},$$
that points to the uniform distribution of critical current density
through the area of the sample and the absence of large scale domain
structure. The main result of the article is shown on fig. \ref{IT}.
One can see that the critical current vs temperature dependence has
a reentrant behaviour at 4.2 K. As it was shown in Refs. \cite{PRL}
and \cite{SFS-06} a reentrant $I_c(T)$ curve gives an evidence of a
transition into the $\pi$-state. However our data do not allow to
specify what is the ground state of the junction (0 or $\pi$) at
given temperature range. To answer this question one should explore
$j_c(d_F)$ dependance in a wide range of $d_F$ in order to estimate
the period $\lambda_{ex}$ as it was done, for example, in Ref.
\cite{SFS-06}.

In conclusion, we have studied the magnetic and transport properties
of $Pd_{0.99}Fe_{0.01}$ thin films and SFS josephson junctions
$Nb-Pd_{0.99}Fe_{0.01}-Nb$. It was found that the temperature
dependance of the junction with ferromagnet layer thickness equal to
38 nm demonstrates the reentrant behaviour. According to earlier
work this is the evidence of the transition of the sample into the
$\pi$-state. Authors are grateful to V.V. Ryazanov and V.A. Oboznov
for helpful discussions and N.S. Stepakov for the assistance during
experiments and sample preparation. We also thank Marat  Gaifullin
from National  Institute  for  Material  Science (Japan) for SQUID
magnetometry measurements. This work has been done under Russian
Foundation for Basic Research grant 05-02-17731-a.

\end{document}